\documentclass[letter,twocolumn]{jpsj3}
\usepackage{txfonts}
\usepackage{bm}
\usepackage{color}

\title{Fractional Quantum Hall Effect in $\bm{n}=0$ Landau Band of Graphene 
with Chern Number Matrix}

\author{Koji Kudo$^1$ and Yasuhiro Hatsugai$^{1,2}$}
\inst{$^1$Graduate School of Pure and Applied Science, University of Tsukuba, Tsukuba 305-8571, Japan\\
$^2$Division of Physics, University of Tsukuba, Tsukuba, Ibaraki 305-8571, Japan}
\abst{
Fully taking into account the honeycomb lattice structure, fractional 
quantum Hall states of graphene are considered by a pseudopotential projected 
into the $n=0$ Landau band. By using chirality as an internal degree of 
freedom, the  Chern number matrices are defined and evaluated numerically. 
Quantum phase transition induced by changing a range of the interaction is
demonstrated that is associated with chirality ferromagnetism. The 
chirality-unpolarized ground state is consistent with the Halperin 331 state of
the bilayer quantum Hall system.
}

\begin{document}

\maketitle
The concept of topological order
\cite{PhysRevB.40.7387,doi:10.1143/JPSJ.73.2604,
doi:10.1143/JPSJ.74.1374} substantially expands our understanding of phases of
matter that cannot be described by the conventional order parameters
associated with symmetry breaking. The quantum Hall effect
\cite{PhysRevLett.45.494,PhysRevLett.48.1559} is a prominent example
of topologically non-trivial quantum phases. The quantized Hall conductance is
expressed as the Chern number
\cite{PhysRevLett.49.405,PhysRevB.31.3372,KOHMOTO1985343}
associated with the Berry connection\cite{Berry84}. Although the integer 
quantum
Hall state can be described by non-interacting electrons, the
electron-electron interaction plays a crucial role in the fractional quantum
Hall (FQH) phase. The characteristics of these correlated quantum states are
well captured by the Laughlin wave function\cite{PhysRevLett.50.1395}. In 
addition,
its excitations are quasiparticles with fractional
charges and fractional statistics\cite{PhysRevLett.53.722}. This FQH effect
is understood as an integer quantum Hall effect of composite fermions as flux
charge composites \cite{PhysRevLett.63.199}. It also provides a consistent 
picture
at an even-denominator Landau level (LL) filling \cite{PhysRevB.47.7312}.

The internal degrees of freedom, such as spin or layer index, bring further
diversity to the FQH phases. The ground state at $\nu=5/2$
\cite{PhysRevLett.59.1776} is described by the Moore-Read
Pfaffian state \cite{1991NuPhB.360..362M} with the excitation obeying
non-Abelian statistics \cite{PhysRevB.54.16864} when the interaction is 
short-range. As for two-component Abelian FQH systems, the Halperin 
$lmn$
state \cite{Halperin:1983zz} is a typical example. This is realized in a
bilayer FQH system at $\nu=1/2$ \cite{PhysRevB.39.1932,PhysRevB.47.4394,
PhysRevLett.68.1383,doi:10.1143/JPSJ.73.2612,
PhysRevB.81.165304,PhysRevB.82.075302}, for example, but its
appearance depends strongly on the system parameters. For 
multi-component systems, the symmetric integer matrix $\bm{K}$
\cite{PhysRevB.46.2290,PhysRevLett.69.953,PhysRevB.42.8133,PhysRevB.43.8337}
provides classification of the FQH phases, which is discussed in relation to
the Chern number matrix \cite{PhysRevLett.91.116802,PhysRevB.95.125134}.

The FQH effect of graphene \cite{nature462_192,nature08582-3,nphys2007,
PhysRevLett.106.046801,Feldman1196} is also an example of the
multi-component FQH systems. The low-energy behavior of electrons in graphene 
is
described by the doubled massless Dirac fermions at $K$ and $K'$ points in the
Brillouin zone. These characteristics give rise to the FQH
phases peculiar to graphene. 
\cite{PhysRevLett.96.256602,PhysRevLett.97.126801,PhysRevB.74.235417,
PhysRevB.75.245440,PhysRevB.77.235426,PAPIC20091056,doi:10.1143/JPSJ.78.104708,
PhysRevLett.105.176802,PhysRevLett.107.176602,PhysRevB.88.115407,
PhysRevB.92.075410}.
Since the $n=0$ LL is a standard lowest LL of the valley polarized Dirac
fermions, the FQH effect of the $n=0$ LL has been discussed similarly with the
SU(2) invariance arising from the valley degree of freedom. The ground states
at the $n=0$ LL filling factor $\nu=1/3$ and $1/2$ are described by the
pseudospin (valley) polarized Laughlin state and pseudospin singlet composite
fermion Fermi sea, respectively.
\cite{PhysRevLett.97.126801,PhysRevB.74.235417}

In this study, the FQH system for the $n=0$ Landau band is investigated
by fully taking into account the honeycomb lattice structure of the
interaction. Short-range
electron-electron interaction of the nearest neighbor (NN) and next-nearest 
neighbor (NNN) is
discussed in this paper by constructing the pseudopotential 
\cite{PhysRevLett.51.605}
projected into the $n=0$ Landau band. The chiral symmetry of the honeycomb 
lattice plays an important role in the many-body problems as well
\cite{PhysRevB.86.205424,PhysRevB.88.195141,1367-2630-15-3-035023}.
The quantum phase transitions associated with the chirality ferromagnetism
occur by changing the interaction range. Since the total pseudospin is not
conserved because of the lattice effects, the SU(2) symmetry discussed by the
continuous approximation is absent. In order to characterize 
the quantum phases topologically, the Chern number matrices specified by the 
chiral basis are constructed numerically. The results are also discussed in
relation to the conventional bilayer quantum Hall system.

Let us begin by introducing the projected fermion operators into
the $n=0$ Landau band. Here, we assume that the system is always
spin-polarized. The kinetic Hamiltonian is written as
\begin{align}
 H_\text{kin}
 =t\sum_{\langle ij\rangle}e^{i\phi_{ij}}c_i^\dagger c_j
 =\bm{c}^\dagger h_\text{kin}\bm{c},
\end{align}
which describes hopping between the NN pairs of sites with
strength $t$. Here, $\bm{c}^\dagger=(\bm{c}^\dagger_\bullet,\bm{c}^\dagger
_\circ)$, $\bm{c}^\dagger_{\bullet(\circ)}=(c^\dagger_{1\bullet(\circ)},\cdots,
c^\dagger_{N_\text{cell}\bullet(\circ)})$ and
$c^\dagger_{i\bullet(\circ)}$ creates a fermion at the sublattice $\bullet
(\circ)$ for unit cell $i$.
The Peierls phase $\phi_{ij}$ is determined such that the sum of the phases
around an elementary hexagon is equal to the magnetic flux $2\pi\phi$ in units 
of the flux quantum $\phi_0=h/e$. In the calculation, the string gauge
\cite{PhysRevLett.83.2246,doi:10.7566/JPSJ.86.103701} is employed,
which enables us to realize the minimum magnetic fluxes that are consistent 
with the lattice periodicity. When the system is put on the $N_x\times N_y$
unit cells with a periodic boundary condition, the magnetic field can be
provided as $\phi=N_\phi/N_\text{cell}\ (N_\phi=1,2,\cdots,N_\text{cell})$,
where $N_\text{cell}=N_xN_y$. Here, $N_\phi$ corresponds to the total magnetic
flux. The lattice model with $\phi=p/q\,$($p,q:$ relatively prime) has $2q$
single-electron bands, where 2 comes from the sublattice degree of freedom.
The number of states per band is obtained as $N_xN_y/q$. For the weak magnetic
field ($\phi\ll1$), $2p$ bands flow into each other around the zero energy, which
form the $n=0$ LL in the large $q$ limit. Thus, in this paper, ``the $n=0$
Landau band'' is defined as a group of these bands, where there are
$(N_xN_y/q)\times2p=2N_\phi$ one-body states. 

Since the honeycomb lattice is bipartite, the Hamiltonian $H_\text{kin}$ has
chiral symmetry. The $2N_\text{cell}\times 2N_\text{cell}$ matrix
$\Gamma=\text{diag}(I_{N_\text{cell}},-I_{N_\text{cell}})$ anticommutes with
the Hamiltonian as $\{h_\text{kin},\Gamma\}=0$ and $\Gamma^2=I_{2N_
\text{cell}}$. If $\psi_k$ is the eigenvector of $h_\text{kin}$ with the energy
$\epsilon_k$, the chiral symmetry guarantees that $\Gamma\psi_k$ is identical
to the one with $-\epsilon_k$. Thus, the chiral operator $\Gamma$ can be
diagonalized within the one-body states of the $n=0$ Landau band. Then, the 
chiral basis can be defined as $\psi=(\psi_+,\psi_-)$,
$\psi_\pm=(\psi_{1,\pm},\cdots,\psi_{N_\phi,\pm})$, and 
$\Gamma\psi_{k,\pm}=\pm\psi_{k,\pm}$. Note that $\psi_{k,+(-)}$ is localized on
the sublattice $\bullet(\circ)$ so that the multiplet can be expressed as 
$
\psi=\bigl(
\begin{smallmatrix}
 \psi_{\bullet} & 0\\ 
 0 & \psi_{\circ}
\end{smallmatrix}\bigr)
$,
where $\psi_{\bullet(\circ)}$ is a proper $N_\text{cell}\times N_\phi$ matrix
\cite{1367-2630-15-3-035023}.

Next, let us consider the two-body interactions written as
\begin{align}
 H_\text{int}=
 \sum_{i<j}\sum_{\sigma=\bullet,\circ}
 V_{ij}^{\sigma\sigma}n_{i\sigma}n_{j\sigma}
 +\sum_{i,j}
 V_{ij}^{\bullet\circ}n_{i\bullet}n_{j\circ},
\end{align}
where $n_{i\bullet(\circ)}=c_{i\bullet(\circ)}^\dagger c_{i\bullet(\circ)}$ and
$V_{ij}^{\sigma\sigma'}$ is the strength of the electron-electron interaction.
In order to construct the pseudopotential, the projected creation-annihilation
operators are defined as $\tilde{\bm{c}}^\dagger=\bm{c}^\dagger P$
\cite{PhysRevB.86.205424,PhysRevB.88.195141,1367-2630-15-3-035023,
doi:10.7566/JPSJ.86.103701}, where $P=\psi\psi^\dagger$. This expression is
simplified by writing $\tilde{\bm{c}}^\dagger_{\bullet(\circ)}
=\bm{c}^\dagger_{\bullet(\circ)}P_{\bullet(\circ)}$ and 
$P_{\bullet(\circ)}=\psi_{\bullet(\circ)}\psi^\dagger_{\bullet(\circ)}$.
Note that these projected operators no longer satisfy the canonical
anticommutation relations ($\{\tilde{c}_i,\tilde{c}^\dagger_j\}=
P_{ij}\neq\delta_{ij}$). By taking into account the ordering of fermions,
the replacement of $\bm{c}^\dagger$, $\bm{c}$ with $\tilde{\bm{c}}^\dagger$,
$\tilde{\bm{c}}$ causes the Hamiltonian to be projected into the $n=0$ Landau
band. Then, the projected Hamiltonian can be defined as
\begin{align}
 \tilde{H}_\text{int}
 &=\sum_{i<j}\sum_{\sigma=\bullet,\circ}
 V_{ij}^{\sigma\sigma}
 \tilde{c}^\dagger_{i\sigma}\tilde{c}^\dagger_{j\sigma}
 \tilde{c}_{j\sigma}\tilde{c}_{i\sigma}
 +\sum_{i,j}
 V_{ij}^{\bullet\circ}
 \tilde{c}^\dagger_{i\bullet}\tilde{c}^\dagger_{j\circ}
 \tilde{c}_{j\circ}\tilde{c}_{i\bullet}
 \label{eq:pro_H}\\
 &=\sum_{klmn}
 (\sum_{\sigma=\bullet,\circ}A^{\sigma\sigma}_{klmn}
 d^\dagger_{k,\,\chi_\sigma}d^\dagger_{l,\,\chi_\sigma}
 d_{m,\,\chi_\sigma}d_{n,\,\chi_\sigma}\nonumber\\
 &\qquad\qquad\qquad\qquad\qquad+A^{\bullet\circ}_{klmn}
 d^\dagger_{k,\,+} d^\dagger_{l,\,-} d_{m,\,-}d_{n,\,+}),
\end{align}
where
$
A^{\sigma\sigma'}_{klmn}
=\sum_{i<j}V_{ij}^{\sigma\sigma'}
(\psi_\sigma)^\ast_{ik}(\psi_{\sigma'})^\ast_{jl}
(\psi_{\sigma'})_{jm}(\psi_\sigma)_{in},
$
$\bm{d}_{+(-)}^\dagger=(d^\dagger_{1,\,+(-)},\cdots,d^\dagger_{N_\phi,\,+(-)})
=\bm{c}_{\bullet(\circ)}^\dagger\psi_{\bullet(\circ)}$, and $\chi_
{\bullet(\circ)}=+(-)$. Here, we choose the strength of the interaction such
that its energy scale is much larger than the energy width of the $n=0$ Landau
band, so that only the interaction term is considered. The projected
Hamiltonian $\tilde{H}_\text{int}$ commutes with the total chirality operator
written as
\begin{align}
 \mathcal{G}=\tilde{\bm{c}}^\dagger\Gamma
\tilde{\bm{c}}=\bm{d}^\dagger_+\bm{d}_+-\bm{d}^\dagger_-\bm{d}_-,
\end{align}
which enables us to classify the $N_\text{e}$-body states by the total 
chirality
$\chi_\text{tot}=-N_\text{e},-N_\text{e}+2,\cdots,N_\text{e}$. Now, the filling
factor is defined as $\nu=N_\text{e}/N_\phi$.

\begin{figure}[!t]
  \begin{center}
   \includegraphics[width=\columnwidth]{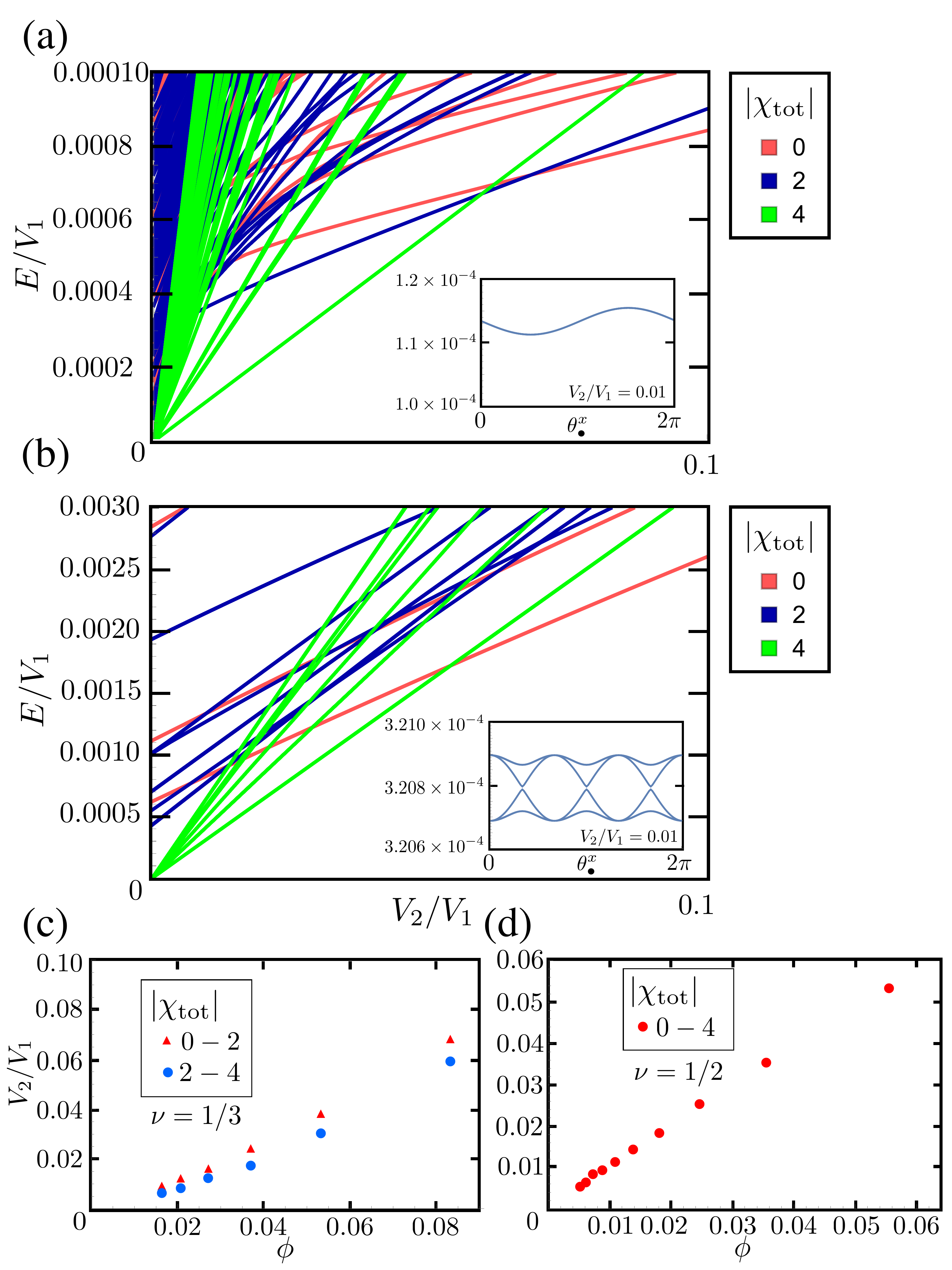}
  \end{center}
 \caption{(Color online) (a, b) Many-body spectrum as a function of the ratio
 $V_2/V_1$ at filling factor (a) $\nu=1/3$ and (b) $\nu=1/2$. 
 The total chiralities are expressed by the line colors.
 The insets show the many-body spectra with 
 $\chi_\text{tot}=N_\text{e}$ for $V_2/V_1=0.01$. The horizontal axis is the 
 twist $\theta^x_\bullet$ ($\theta^y_\bullet=0$). (See Eq.~(\ref{eq:twist}).) 
 Note that the level crossing in the inset of (b) is not exact.
 (c, d) Dependence of the phase transition point $V_2/V_1$ on 
 magnetic flux $\phi$ at (c) $\nu=1/3$ and (d) $\nu=1/2$.
 The transition point denoted as $a$-$b$ indicates the transition between 
 $\chi_\text{tot}=a$ and $b$.}
 \label{fig:PT1_3}
\end{figure}

Hereafter, we consider the electron-electron interaction between
NN and NNN pairs, 
\begin{align}
 \tilde{H}_\text{int}
 =V_1\sum_{\langle ij\rangle}
 \tilde{c}^\dagger_i\tilde{c}^\dagger_j\tilde{c}_j\tilde{c}_i
 +V_2\sum_{\langle\langle ij\rangle\rangle}
 \tilde{c}^\dagger_i\tilde{c}^\dagger_j\tilde{c}_j\tilde{c}_i,
\end{align}
where $V_1$ and $V_2$ are the strength of the interaction. Note that the
interaction $V_1$ ($V_2$) contributes $V^{\bullet\circ}_{ij}$ ($V^{\bullet
\bullet}_{ij}$ and $V^{\circ\circ}_{ij}$). Figures \ref{fig:PT1_3} (a) and (b)
show the $V_2/V_1$ dependence of the four-electron energy obtained by the 
exact diagonalization at $\nu=1/3$ and $1/2$.
Since the exact topological degeneracy due to the 
center-of-mass translation \cite{PhysRevLett.55.2095} is given by the Landau 
gauge in a finite system, we choose a system of $12\times12$ size for 
$\phi=1/12$ as shown in Fig.~\ref{fig:PT1_3} (a) at $\nu=1/3$. For example, the
chirality-polarized ground state multiplet is exactly three-fold degenerated. 
(See the inset in Fig.~\ref{fig:PT1_3} (a).) As for the system at $\nu=1/2$, to
be compatible to the Dirac cones at $K$ and $K'$ points of graphene, the system
size should be $3n\times3n$ ($n=1,2,\cdots$). Then, we choose the magnetic flux
as $\phi=8/(12\times12)=1/18$ using the string gauge. In this case, the 
topological degeneracy is not exact. However, the energy separation of the 
entangled states within the ground state multiplet is very small. For example, 
it is less than 0.003 times the energy gap for the six-fold ground state 
multiplet with $\chi_\text{tot}=N_\text{e}$. 
(See the inset in Fig.~\ref{fig:PT1_3} (b).)
Note that the magnetic length $l_\text{B}=\sqrt{\hbar/eB}=\sqrt{S/2\pi\phi}$, 
where $S=3\sqrt{3}a^2/2$ is the area of the elementary hexagon with lattice
constant $a$, is approximately $2.2a$ and $2.7a$ for $\phi=1/12$ and $1/18$,
respectively.
In the case of $V_1\neq V_2=0$, that is, only the bipartite interactions, 
many-body states with $|\chi_\text{tot}|=N_\text{e}$ should be eigenstates with
zero eigenvalues. Generally, the system at $\nu<1$ 
provides $2\times{}_{N_\phi}C_{N_\text{e}}$-fold degenerate ground states 
(2 comes from the sign of chirality), which are lifted by the infinitesimal 
$V_2$. Further, an increase in $V_2$ leads the phase transition from $|\chi
_\text{tot}|=N_\text{e}$ to $0$ since the NNN interactions act between the same
sublattices. In Fig.~\ref{fig:PT1_3} (c) and (d), the ratio
$V_2/V_1$ at the phase transition points is plotted against the magnetic flux
$\phi$ at $\nu=1/3$ and $1/2$. 
The results suggest that the
strong short-range interaction compared with the magnetic flux ($\phi\ll1$)
favors the unpolarized chirality unless $V_2$ is not vanishing.

Since the pseudopotential is constructed on the basis of the honeycomb lattice 
model,
only the $z$-component of the pseudospin, $\chi_\text{tot}$, is conserved. The
SU(2) symmetry arising from the chirality is absent in contrast to the cases
in the continuum limit \cite{PhysRevLett.97.126801,PhysRevB.74.235417}. Then,
in order to investigate the internal topological structure of
the many-body states, we evaluate the Chern number matrices associated with the
chirality. First, we investigate the twisted boundary condition
\cite{PhysRevB.31.3372},
$c^\dagger_{n_x+N_x,n_y}=e^{i\theta^x}c^\dagger_{n_x,n_y}$ and
$c^\dagger_{n_x,n_y+N_y}=e^{i\theta^y}c^\dagger_{n_x,n_y}$, where $n_x$ and
$n_y$ are the labels of the unit cell for the $x$ and $y$ directions. Since the
chiral symmetry remains in the kinetic Hamiltonian $H_\text{kin}(\theta^x,
\theta^y)$, the connection between chirality and sublattice is preserved. 

Let us consider each projected fermion operator $\tilde{\bm{c}}_\bullet$
and $\tilde{\bm{c}}_\circ$ with different twisted boundary conditions,
\begin{align}
 \tilde{\bm{c}}_{\bullet(\circ)}
 =\tilde{\bm{c}}_{\bullet(\circ)}(\theta_{\bullet(\circ)}),
 \label{eq:twist}
\end{align}
where $\theta_{\bullet(\circ)}=(\theta_{\bullet(\circ)}^x,\theta_{\bullet
(\circ)}^y)$. The projected Hamiltonian $\tilde{H}_\text{int}(\theta_\bullet,
\theta_\circ)$ can be defined by replacing $\tilde{\bm{c}}_{\bullet(\circ)}$
with $\tilde{\bm{c}}_{\bullet(\circ)}(\theta_{\bullet(\circ)})$ in
Eq.~(\ref{eq:pro_H}). Note that this Hamiltonian can be written by the fermion
operators $\bm{d}^\dagger_{+(-)}(\theta_{\bullet(\circ)})=
\bm{c}^\dagger_{\bullet(\circ)}\psi_{\bullet(\circ)}(\theta_{\bullet(\circ)})$,
which satisfy the canonical anticommutation relations 
$\{d_{i,\,+}(\theta_\bullet),d_{j,\,-}(\theta_\circ)\}
=\{d^\dagger_{i,\,+}(\theta_\bullet),d_{j,\,-}(\theta_\circ)\}=0$
for any $\theta_{\bullet(\circ)}$.

Now, let us further define the non-Abelian Berry connection and curvature
\cite{Berry84,doi:10.1143/JPSJ.73.2604,doi:10.1143/JPSJ.74.1374}
of the $m$-fold ground state multiplet $\Phi=(|G_1\rangle,\cdots,
|G_m\rangle)$ by selecting the sublattices in each direction $x,y$ as
\begin{align}
 &\bm{A}_{\sigma_x\sigma_y}
 =\Phi^\dagger d\Phi,\ 
 d=\sum_{\mu=x,y}d\theta_{\sigma_\mu}^\mu\frac{\partial}{\partial \theta_{\sigma_\mu}^\mu},\\
 &\bm{F}_{\sigma_x\sigma_y}
 =d\bm{A}_{\sigma_x\sigma_y}+\bm{A}_{\sigma_x\sigma_y}^2,
\end{align}
where $\sigma_\mu=\bullet,\circ$, and the two parameters except for
$\theta^x_{\sigma_x}$ and $\theta^y_{\sigma_y}$ are fixed to 0. Then, the Chern
number matrix is defined as
\begin{align}
 &\bm{C}
 =\left(
 \begin{array}{cc}
  C_{\bullet\bullet}& C_{\bullet\circ} \\
  C_{\circ\bullet}& C_{\circ\circ} \\
 \end{array}
 \right),\ 
 C_{\sigma_x\sigma_y}
 =\frac{1}{2\pi i}\int_{T^2}
 \text{Tr}\,\bm{F}_{\sigma_x\sigma_y}.
\end{align}
The element is evaluated as 
\cite{doi:10.1143/JPSJ.74.1674,doi:10.7566/JPSJ.86.103701}
$C_{\sigma_x\sigma_y}
=\frac{1}{2\pi i}\sum_{\theta_{\sigma}}\tilde{F}_{\sigma_x\sigma_y}
(\theta_{\sigma})$ numerically, where 
$\tilde{F}_{\sigma_x\sigma_y}(\theta_\sigma)
 =\text{Log}\,
 [U_{\sigma_x}^x(\theta_\sigma)
 U_{\sigma_y}^y(\theta_\sigma +\Delta_{\sigma_x}^x)
 U_{\sigma_x}^x(\theta_\sigma+\Delta_{\sigma_y}^y)^{-1}
 U_{\sigma_y}^y(\theta_\sigma)^{-1}]
$,
$
U_{\bullet(\circ)}^\mu(\theta_\sigma)
 =\det[\Phi^\dagger(\theta_\sigma)
 \Phi(\theta_\sigma+\Delta_{\bullet(\circ)}^\mu)]
 /|\det[\Phi(\theta_\sigma)
 \Phi(\theta_\sigma+\Delta_{\bullet(\circ)}^\mu)]|
$,
$\theta_\sigma=(\theta_{\sigma_x}^x,\theta_{\sigma_y}^y)$, and
$\Delta_{\bullet(\circ)}^\mu$ represents the lattice displacement in the
direction $\mu=x,y$ for the sublattice $\bullet(\circ)$.

In order to construct the ground state multiplet $\Phi(\theta_\sigma)$
numerically, a basis of $N_\text{e}$-electron states classified by the total
chirality $\chi_\text{tot}$ is defined as
$\Psi(\theta_\sigma)
=(|\Psi_1(\theta_\sigma)\rangle,\cdots, |\Psi_{N_D}(\theta_\sigma)\rangle)
$.
Here, $N_D$ is the dimension of the Hilbert space, and
\begin{align}
 |\Psi_i(\theta_\sigma)\rangle
 =\left(\prod_{n\in P_{i,\,+}}d_{n,\,+}^\dagger(\theta_\sigma)\right)
 \left(\prod_{n\in P_{i,\,-}}d_{n,\,-}^\dagger(\theta_\sigma)\right)
 |0\rangle,
\end{align}
where $P_{i,\,\pm}$ is one of the possible ways to occupy $N_\phi$ states by
$N_\pm$ electrons, and $N_\pm=(N_\text{e}\pm\chi_\text{tot})/2$. Then, the
ground state multiplet is expressed as $\Phi=\Psi u_G$, where $u_G=
(\bm{u}_{G_1},\cdots,\bm{u}_{G_m})$, and $\bm{u}_{G_i}$ is the eigenvector of
$\Psi^\dagger\tilde{H}_\text{int}\Psi$. We have
$
\Phi^\dagger(\theta_\sigma)\Phi(\theta_\sigma
+\Delta_{\bullet(\circ)}^\mu)=u_G^\dagger(\theta_\sigma)O(\theta_\sigma,
\Delta_{\bullet(\circ)}^\mu)u_G(\theta_\sigma+\Delta_{\bullet(\circ)}^\mu)
$ and
\begin{align}
 O_{ij}(\theta_\sigma,\Delta_{\bullet(\circ)}^\mu)
 =&\langle\Psi_i(\theta_\sigma)|\Psi_j(\theta_\sigma
 +\Delta_{\bullet(\circ)}^\mu)\rangle\nonumber\\
 =&\delta_{P_{i,\,-(+)}P_{j,\,-(+)}}
 \langle0|
 \left(
 \prod_{n\in P_{i,\,+(-)}}d_{n,\,+(-)}^\dagger(\theta_\sigma)
 \right)^\dagger\nonumber\\
 &\qquad\qquad\qquad
 \left(\prod_{n\in P_{j,\,+(-)}}d_{n,\,+(-)}^\dagger(\theta_\sigma
 +\Delta_{\bullet(\circ)}^\mu)\right)|0\rangle\nonumber\\
 =&\delta_{P_{i,\,-(+)}P_{j,\,-(+)}}
 \det[\tilde{\psi}_{P_{i,\,+(-)}}^\dagger
 (\theta_\sigma)
 \tilde{\psi}_{P_{j,\,+(-)}}
 (\theta_\sigma+\Delta_{\bullet(\circ)}^\mu)],\nonumber
\end{align}
where
$
\tilde{\psi}_{P_{i,\,+(-)}}
=(\psi_{\alpha_1,\,\bullet(\circ)},\cdots,
\psi_{\alpha_{N_\text{e}},\,\bullet(\circ)})
$,
$\psi_{i,\,\bullet(\circ)}$ is the $i$-th column vector of $\psi_{\bullet
(\circ)}$, and $P_{i,\,+(-)}=\{\alpha_1\cdots,\alpha_{N_\text{e}}\}$.

We first focus on the chirality-polarized states with $\chi_\text{tot}=
N_\text{e}$ for $V_2\neq0$. Since the polarized many-body states occupy only
the sublattice $\bullet$, the staggered sublattice potential written as 
$
\tilde{H}_\text{site}
=-M\sum_{i}\tilde{c}_{i\bullet}^\dagger\tilde{c}_{i\bullet}
+M\sum_{i}\tilde{c}_{i\circ}^\dagger\tilde{c}_{i\circ}
=-M\mathcal{G}\ 
(M>0)$ stabilizes the polarized many-body states. Now, the ground
state multiplet is defined as a group of lowest-energy many-body states that is
separated from the other excited states. Numerically obtained ground
states at $\nu=1/3$ and $1/2$ are similar to those
provided by the pseudopotentials projected into the lowest Landau band
\cite{doi:10.7566/JPSJ.86.103701}.
At $\nu=1/3$, the ground state is always three-fold degenerated 
irrespective of the number of electrons. This three-fold ground state multiplet
is gapped in the thermodynamic limit similar to the
Ref.~\citen{doi:10.7566/JPSJ.86.103701}. In addition, its Chern number
$C_{\bullet\bullet}$ is 1. 
This implies that the ground state is the lattice analogue of the Laughlin 
state.
On the other hand, for 
the $\nu=1/2$ case, the degeneracy of the ground state has no such universal 
feature, and there is no sign of a finite energy gap from its scaling. This 
is also consistent with the composite fermion Fermi sea
\cite{doi:10.7566/JPSJ.86.103701}.

\begin{figure}[tb]
  \begin{center}
   \includegraphics[width=\columnwidth]{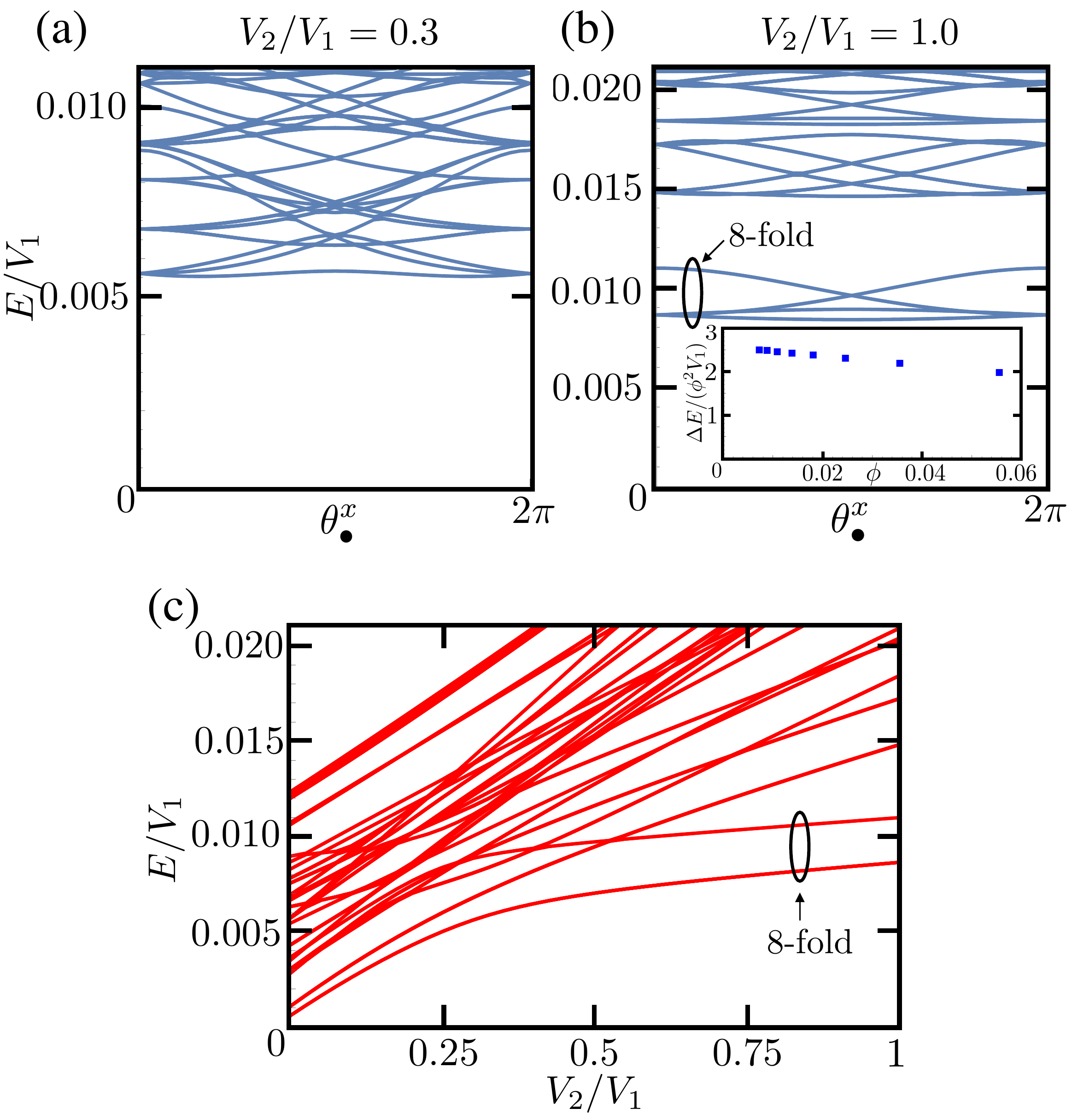}
  \end{center}
 \caption{(Color online) Many-body spectrum with $\chi_\text{tot}=0$ at
 $\nu=1/2$  as a function of (a, b) the twist
 $\theta^x_\bullet$ and (c) the ratio of the interaction $V_2/V_1$.
 (a, b) The remaining parameters are fixed at $\theta^y_\bullet=\theta^x_\circ
 =\theta^y_\circ=0$. The inset in (b) shows the scaled energy gaps as a
 function of the magnetic flux $\phi$.}
 \label{fig:331_twist}
\end{figure}

Next, let us consider the many-body states with $\chi_\text{tot}=0$, which
occupy the same number of sublattices $\bullet$ and $\circ$. Here, we  
focus on the $\nu=1/2$ state. Figure~\ref{fig:331_twist} (a) and (b) shows the
$\theta_x^\bullet$ dependence of the many-body spectrum at $V_2/V_1=0.3$ and 
$1.0$, respectively. In Fig.~\ref{fig:331_twist} (a), the ground state mixes 
with
higher states with the change in boundary conditions, and the ground state
multiplet is not well-defined. On the other hand, in Fig.~\ref{fig:331_twist} 
(b), eight low-energy states are entangled and do not mix with excited states. 
In Fig.~\ref{fig:331_twist} (c), the many-body spectrum with 
$\chi_\text{tot}=0$ for $N_\text{e}=4$ and $\phi=1/18$ is plotted as a function
of $V_2/V_1$. 
The behavior of the ground state in the spectral flow changes at 
$V_2/V_1\simeq0.53$. 
Note that the eight-fold ground state multiplet is always gapped for $V_2/V_1
\gtrsim0.53$ as long as $V_1\neq0$ (For $V_1=0$, the energy gap of the 
eight-fold
ground state multiplet vanishes, and the two decoupled $\nu/2+\nu/2$ states
should be the ground states). In the inset of the Fig.~\ref{fig:331_twist} (b),
the energy gaps  $\Delta E=E_{9}-E_1$ at $V_2/V_1=1$ with periodic boundary
conditions are plotted as a function of the magnetic flux $\phi$, where $E_i$
is the $i$-th eigenvalue of $\tilde{H}_\text{int}$ with $\chi_\text{tot}=0$. 
The result indicates that the scaling low $\Delta E\propto\phi^2$ is roughly
valid in the wide range of $\phi$. Since the electron density is obtained as
$\rho=N_\text{e}/(2N_\text{cell})=\phi\nu/2$, we have $\Delta E\propto\rho^2$,
which means that the excitations are local.

Following the argument of the Ref.~\citen{PhysRevB.95.125134}, the obtained 
Chern number matrix $\bm{C}$ suggests the $\bm{K}$-matrix as $\bm{K}=(\bm{C}/m)
^{-1}$, where $m$ is the degeneracy.
The matrices defined by the eight-fold ground state at $\nu=1/2$ are 
numerically observed as
\begin{align}
 &\bm{C}
 =\left(
 \begin{array}{cc}
  3 & -1\\
  -1 & 3
 \end{array}
 \right),\ 
 \bm{K}
 =\left(
 \begin{array}{cc}
  3 & 1\\
  1 & 3
 \end{array}
 \right).
 \label{eq:K_matrix}
\end{align}

As for a conventional bilayer FQH system with finite 
interlayer separation $d$, the pseudopotential does not have the SU(2) symmetry
arising from the pseudospin. However, in the limit $d\rightarrow\infty$ or 
$d=0$, the system is exceptionally SU(2) invariant. Therefore, the ground 
states of the $\nu=1/2$ case in these two limits are described by the 
composite 
fermion Fermi sea. The former is the two decoupled states and the latter is the
pseudospin-singlet state.
\cite{PhysRevB.64.085313}
On the other hand, the Halperin 331 state, which does not have the SU(2) 
symmetry, is realized as the ground state in the intermediate separation.
\cite{PhysRevB.39.1932,PhysRevB.47.4394,PhysRevLett.68.1383}
The FQH system for the $n=0$ LL of graphene in the continuum limit 
corresponds to the conventional bilayer FQH system with $d=0$, where the 
Halperin 331 state should not be observed because there is SU(2) symmetry 
in the system.

In contrast, the NN and NNN interactions in 
the honeycomb lattice act between the different and same sublattices 
(chiralities). 
Therefore, by connecting the parameter $V_1$ and $V_2$ with the interlayer 
and intralayer interactions, the FQH system of the $n=0$ Landau band 
corresponds to the conventional bilayer quantum Hall system 
with $d\neq0$; the 
chirality $\pm1$ plays the role of the two-layer index. Note that the only 
$z$-component of the pseudospin is conserved in both systems.
Thus, it is natural to expect that the analogue of the Halperin
331 state is realized for the $n=0$ Landau band when the ratio $V_2/V_1$ is on
the order of unity. We confirmed this scenario for the four-electron system 
in terms of the Chern number matrix as Eq.~(\ref{eq:K_matrix}).

The chirality-unpolarized ground state at $\nu=1/3$ is also considered in terms
of the Chern number matrix. In the conventional bilayer system at $\nu=1/3$,
for example, the Halperin 551 state is one of the candidates of the gapped 
ground state\cite{PhysRevB.39.1932,PhysRevB.64.085313}. 
However, the results at $\nu=1/3$ are not systematic, and no clear picture is 
obtained in contrast to the $\nu=1/2$ case. They will be discussed later
elsewhere.

To summarize, we have constructed the Chern number matrix in association with
the chiral basis by using the Hamiltonian projected into the $n=0$ 
Landau band of the honeycomb lattice. Modifying the interaction range induces 
the quantum phase transitions associated with the chirality ferromagnetism.
When the NN interactions are sufficiently strong, the ground 
states at $\nu=1/3$ and $1/2$ are chirality-polarized, and consistent with the 
Laughlin state and the composite fermion Fermi sea, respectively. On the 
other hand, an increase in the strength of the NNN interaction
leads the ground state to chirality-unpolarize. The obtained Chern 
number matrix indicates that the unpolarized ground state for large $V_2$ at 
$\nu=1/2$ is consistent with the Halperin 331 state.
\begin{acknowledgments}
This work is partly supported by Grants-in-Aid for Scientific Research, 
(KAKENHI), Grant numbers 17H06138, 16K13845, and 25107005.
\end{acknowledgments}

\bibliographystyle{jpsj}
\bibliography{citation}

\begin{thebibliography}{10}

\bibitem{PhysRevB.40.7387}
X.~G. Wen: Phys. Rev. B {\bfseries 40} (1989) 7387.

\bibitem{doi:10.1143/JPSJ.73.2604}
Y.~Hatsugai: Journal of the Physical Society of Japan {\bfseries 73} (2004)
  2604.

\bibitem{doi:10.1143/JPSJ.74.1374}
Y.~Hatsugai: Journal of the Physical Society of Japan {\bfseries 74} (2005)
  1374.

\bibitem{PhysRevLett.45.494}
K.~v. Klitzing, G.~Dorda, and M.~Pepper: Phys. Rev. Lett. {\bfseries 45} (1980)
  494.

\bibitem{PhysRevLett.48.1559}
D.~C. Tsui, H.~L. Stormer, and A.~C. Gossard: Phys. Rev. Lett. {\bfseries 48}
  (1982) 1559.

\bibitem{PhysRevLett.49.405}
D.~J. Thouless, M.~Kohmoto, M.~P. Nightingale, and M.~den Nijs: Phys. Rev.
  Lett. {\bfseries 49} (1982) 405.

\bibitem{PhysRevB.31.3372}
Q.~Niu, D.~J. Thouless, and Y.-S. Wu: Phys. Rev. B {\bfseries 31} (1985) 3372.

\bibitem{KOHMOTO1985343}
M.~Kohmoto: Annals of Physics {\bfseries 160} (1985) 343 .

\bibitem{Berry84}
M.~V. Berry: Proceedings of the Royal Society of London. A. Mathematical and
  Physical Sciences {\bfseries 392} (1984) 45.

\bibitem{PhysRevLett.50.1395}
R.~B. Laughlin: Phys. Rev. Lett. {\bfseries 50} (1983) 1395.

\bibitem{PhysRevLett.53.722}
D.~Arovas, J.~R. Schrieffer, and F.~Wilczek: Phys. Rev. Lett. {\bfseries 53}
  (1984) 722.

\bibitem{PhysRevLett.63.199}
J.~K. Jain: Phys. Rev. Lett. {\bfseries 63} (1989) 199.

\bibitem{PhysRevB.47.7312}
B.~I. Halperin, P.~A. Lee, and N.~Read: Phys. Rev. B {\bfseries 47} (1993)
  7312.

\bibitem{PhysRevLett.59.1776}
R.~Willett, J.~P. Eisenstein, H.~L. St\"ormer, D.~C. Tsui, A.~C. Gossard, and
  J.~H. English: Phys. Rev. Lett. {\bfseries 59} (1987) 1776.

\bibitem{1991NuPhB.360..362M}
G.~{Moore} and N.~{Read}: Nuclear Physics B {\bfseries 360} (1991) 362.

\bibitem{PhysRevB.54.16864}
N.~Read and E.~Rezayi: Phys. Rev. B {\bfseries 54} (1996) 16864.

\bibitem{Halperin:1983zz}
B.~I. Halperin: Helv. Phys. Acta {\bfseries 56} (1983) 75.

\bibitem{PhysRevB.39.1932}
D.~Yoshioka, A.~H. MacDonald, and S.~M. Girvin: Phys. Rev. B {\bfseries 39}
  (1989) 1932.

\bibitem{PhysRevB.47.4394}
S.~He, S.~Das~Sarma, and X.~C. Xie: Phys. Rev. B {\bfseries 47} (1993) 4394.

\bibitem{PhysRevLett.68.1383}
J.~P. Eisenstein, G.~S. Boebinger, L.~N. Pfeiffer, K.~W. West, and S.~He: Phys.
  Rev. Lett. {\bfseries 68} (1992) 1383.

\bibitem{doi:10.1143/JPSJ.73.2612}
K.~Nomura and D.~Yoshioka: Journal of the Physical Society of Japan {\bfseries
  73} (2004) 2612.

\bibitem{PhysRevB.81.165304}
M.~R. Peterson and S.~Das~Sarma: Phys. Rev. B {\bfseries 81} (2010) 165304.

\bibitem{PhysRevB.82.075302}
Z.~Papi\ifmmode~\acute{c}\else \'{c}\fi{}, M.~O. Goerbig, N.~Regnault, and
  M.~V. Milovanovi\ifmmode~\acute{c}\else \'{c}\fi{}: Phys. Rev. B {\bfseries
  82} (2010) 075302.

\bibitem{PhysRevB.46.2290}
X.~G. Wen and A.~Zee: Phys. Rev. B {\bfseries 46} (1992) 2290.

\bibitem{PhysRevLett.69.953}
X.~G. Wen and A.~Zee: Phys. Rev. Lett. {\bfseries 69} (1992) 953.

\bibitem{PhysRevB.42.8133}
B.~Blok and X.~G. Wen: Phys. Rev. B {\bfseries 42} (1990) 8133.

\bibitem{PhysRevB.43.8337}
B.~Blok and X.~G. Wen: Phys. Rev. B {\bfseries 43} (1991) 8337.

\bibitem{PhysRevLett.91.116802}
D.~N. Sheng, L.~Balents, and Z.~Wang: Phys. Rev. Lett. {\bfseries 91} (2003)
  116802.

\bibitem{PhysRevB.95.125134}
T.-S. Zeng, W.~Zhu, and D.~N. Sheng: Phys. Rev. B {\bfseries 95} (2017) 125134.

\bibitem{nature462_192}
X.~Du, I.~Skachko, F.~Duerr, A.~Luican, and E.~Y. Andrei: Nature {\bfseries
  462} (2009) 192 EP .

\bibitem{nature08582-3}
K.~I. Bolotin, F.~Ghahari, M.~D. Shulman, H.~L. Stormer, and P.~Kim: Nature
  {\bfseries 462} (2009) 196 EP .

\bibitem{nphys2007}
C.~R. Dean, A.~F. Young, P.~Cadden-Zimansky, L.~Wang, H.~Ren, K.~Watanabe,
  T.~Taniguchi, P.~Kim, J.~Hone, and K.~L. Shepard: Nature Physics {\bfseries
  7} (2011) 693 EP .

\bibitem{PhysRevLett.106.046801}
F.~Ghahari, Y.~Zhao, P.~Cadden-Zimansky, K.~Bolotin, and P.~Kim: Phys. Rev.
  Lett. {\bfseries 106} (2011) 046801.

\bibitem{Feldman1196}
B.~E. Feldman, B.~Krauss, J.~H. Smet, and A.~Yacoby: Science {\bfseries 337}
  (2012) 1196.

\bibitem{PhysRevLett.96.256602}
K.~Nomura and A.~H. MacDonald: Phys. Rev. Lett. {\bfseries 96} (2006) 256602.

\bibitem{PhysRevLett.97.126801}
V.~M. Apalkov and T.~Chakraborty: Phys. Rev. Lett. {\bfseries 97} (2006)
  126801.

\bibitem{PhysRevB.74.235417}
C.~T\ifmmode~\mbox{\H{o}}\else \H{o}\fi{}ke, P.~E. Lammert, V.~H. Crespi, and
  J.~K. Jain: Phys. Rev. B {\bfseries 74} (2006) 235417.

\bibitem{PhysRevB.75.245440}
C.~T\ifmmode~\mbox{\H{o}}\else \H{o}\fi{}ke and J.~K. Jain: Phys. Rev. B
  {\bfseries 75} (2007) 245440.

\bibitem{PhysRevB.77.235426}
N.~Shibata and K.~Nomura: Phys. Rev. B {\bfseries 77} (2008) 235426.

\bibitem{PAPIC20091056}
Z.~Papić, M.~Goerbig, and N.~Regnault: Solid State Communications {\bfseries
  149} (2009) 1056 .

\bibitem{doi:10.1143/JPSJ.78.104708}
N.~Shibata and K.~Nomura: Journal of the Physical Society of Japan {\bfseries
  78} (2009) 104708.

\bibitem{PhysRevLett.105.176802}
Z.~Papi\ifmmode~\acute{c}\else \'{c}\fi{}, M.~O. Goerbig, and N.~Regnault:
  Phys. Rev. Lett. {\bfseries 105} (2010) 176802.

\bibitem{PhysRevLett.107.176602}
Z.~Papi\ifmmode~\acute{c}\else \'{c}\fi{}, R.~Thomale, and D.~A. Abanin: Phys.
  Rev. Lett. {\bfseries 107} (2011) 176602.

\bibitem{PhysRevB.88.115407}
D.~A. Abanin, B.~E. Feldman, A.~Yacoby, and B.~I. Halperin: Phys. Rev. B
  {\bfseries 88} (2013) 115407.

\bibitem{PhysRevB.92.075410}
A.~C. Balram, C.~T\ifmmode~\mbox{\H{o}}\else \H{o}\fi{}ke, A.~W\'ojs, and J.~K.
  Jain: Phys. Rev. B {\bfseries 92} (2015) 075410.

\bibitem{PhysRevLett.51.605}
F.~D.~M. Haldane: Phys. Rev. Lett. {\bfseries 51} (1983) 605.

\bibitem{PhysRevB.86.205424}
Y.~Hamamoto, H.~Aoki, and Y.~Hatsugai: Phys. Rev. B {\bfseries 86} (2012)
  205424.

\bibitem{PhysRevB.88.195141}
Y.~Hamamoto, T.~Kawarabayashi, H.~Aoki, and Y.~Hatsugai: Phys. Rev. B
  {\bfseries 88} (2013) 195141.

\bibitem{1367-2630-15-3-035023}
Y.~Hatsugai, T.~Morimoto, T.~Kawarabayashi, Y.~Hamamoto, and H.~Aoki: New
  Journal of Physics {\bfseries 15} (2013) 035023.

\bibitem{PhysRevLett.83.2246}
Y.~Hatsugai, K.~Ishibashi, and Y.~Morita: Phys. Rev. Lett. {\bfseries 83}
  (1999) 2246.

\bibitem{doi:10.7566/JPSJ.86.103701}
K.~Kudo, T.~Kariyado, and Y.~Hatsugai: Journal of the Physical Society of Japan
  {\bfseries 86} (2017) 103701.

\bibitem{PhysRevLett.55.2095}
F.~D.~M. Haldane: Phys. Rev. Lett. {\bfseries 55} (1985) 2095.

\bibitem{doi:10.1143/JPSJ.74.1674}
T.~Fukui, Y.~Hatsugai, and H.~Suzuki: Journal of the Physical Society of Japan
  {\bfseries 74} (2005) 1674.

\bibitem{PhysRevB.64.085313}
V.~W. Scarola and J.~K. Jain: Phys. Rev. B {\bfseries 64} (2001) 085313.

\end{thebibliography}
\end{document}